\documentclass[fleqn,12pt,twoside]{article}
\usepackage{espcrc1}

\usepackage{graphicx}
\usepackage[figuresright]{rotating}

\newcommand{\AmS}{{\protect\the\textfont2
  A\kern-.1667em\lower.5ex\hbox{M}\kern-.125emS}}

\title{ Phenomenological Equations of State for $SU(N)$
Gauge Theories }

\author{Peter N. Meisinger 
	\address[stupidaddressmark] {Dept. of Physics, 
        Washington University, \\ 
        St. Louis, MO 63130 USA}%
        \thanks{We gratefully acknowledge the support of the U.S. Dept. of
		Energy under DOE DE-FG02-91ER40628},
        Travis R. Miller \addressmark,
        Michael C. Ogilvie \addressmark[stupidaddressmark]}
       
\begin{document}

\maketitle 

\begin{abstract}
Two phenomenological models describing an $SU(N)$ gluon
plasma are presented using the eigenvalues of the Polyakov
loop as the order parameters of the deconfinement transition.
Each model has a
single free parameter and
exhibits behavior similar to lattice simulations over
the range $T_{d}-5T_{d}$.
The $N=2$ deconfinement transition is second order in both
models, while $N=3$,$4$, and $5$ are first order. Both models
appear to have a smooth large-$N\,\ $limit.
The confined
phase is characterized by a mutual repulsion of Polyakov loop
eigenvalues that makes the Polyakov loop expectation value zero. 
The motion of the eigenvalues
is responsible for the approach to the blackbody limit over the
range $T_{d}-5T_{d}$.

\end{abstract}

\maketitle

\section{Models A and B}
\label{sec:modelab}

We have developed\cite{Meisinger:2001cq}
two simple models for
the $SU(N)$ gluon plasma equation of state,
both based on the use of the eigenvalues of
the Polyakov loop $P$ as the natural order parameters of the
deconfinement transition in pure gauge theories
\cite{Pisarski:2000eq}.
The first model is obtained by adding a mass $M$
to the gauge bosons, and working with the high temperature expansion of the
resultant free energy to order $M^2T^2$.
The result is
\begin{equation}
f_{A}\left( P \right) =
T^{4}F_4(P )- M^2 T^2 F_2(P )
\end{equation}
where $F_4$ and $F_2$ are 
simple functions of $P$\cite{Meisinger:2001fi}.
The first term, $T^{4}F_4(P)$,
is precisely the one-loop perturbative result.
Our second model is obtained by supposing
that there is a natural scale $R$ in position space over
which color neutrality is enforced.
This leads to 
\begin{equation}
f_{B}\left( P \right) =T^4 F_4\left( P \right) -\frac{T}
{ R^{3}}\ln \left[ J\left( P \right) \right] 
\end{equation}
where $J\left( P \right) $ is the Jacobian associated with Haar
measure on $SU(N)$.
In both models, the $T^{4}$ term, which favors the deconfined
phase, dominates for large $T$.
The other term,
which favors the confined phase, dominates for small T.
Confinement is naturally obtained from a uniform distribution of
eigenvalues around the unit circle, constrained by the unitary of $P$.
In a pure gauge theory,
below the deconfinement temperature $T_{d}$,
the eigenvalues are frozen in this uniform distribution. 
The deconfinement phase transition results from a
conflict between the two terms. 

\section{$SU(N)$ Thermodynamics for $N=2-5$%
}
\label{sec:thermo}

In both models,
the deconfining transition is second order
for $SU(2)$ and first order for $SU(3)$,
$SU(4)$ and $SU(5)$.
Both models appear to 
quickly approach a large-$N$ limit.
Figures 1-2 show the pressure $p$ 
and the dimensionless interaction measure
$\Delta =(\varepsilon -3p)/T^{4}$ for the
case of $SU(3)$ compared with
results from lattice simulations\cite{Boyd:1996bx}.
Since both models involve only a single dimensional parameter,
these graphs have no free parameters.
Using a transition temperature of $270\,MeV$ for
pure $SU(3)$ gauge theory gives
the plausible values
$M=596\,MeV$ and $R = 1\,$fermi.
It is clear that by allowing the parameters $M$ and $%
R $ to depend on the temperature, a better fit to lattice data can be
obtained at the cost of introducing additional phenomenological parameters.    
Both models show power law
behavior in $\Delta $ for sufficiently large $T$, consistent with $\Delta
\propto 1/T^{2}$, compatible with
$SU(3)$ lattice data.
In comparison,
the Bag model\cite{Cleymans:1986wb},
predicts a $1/T^{4}$ behavior for $\Delta $,
which is ruled out by lattice results.
Using the information contained in the free energy as
a function of $P$,
non-equilibrium as well as equilibrium
behavior can be explored.
For example, in $SU(3)$,
the range of temperatures over which
metastable behavior occurs can be easily determined.

\begin{figure}[htb]
\vspace{-0.4in}
\begin{minipage}[t]{80mm}
\includegraphics[width=3in]{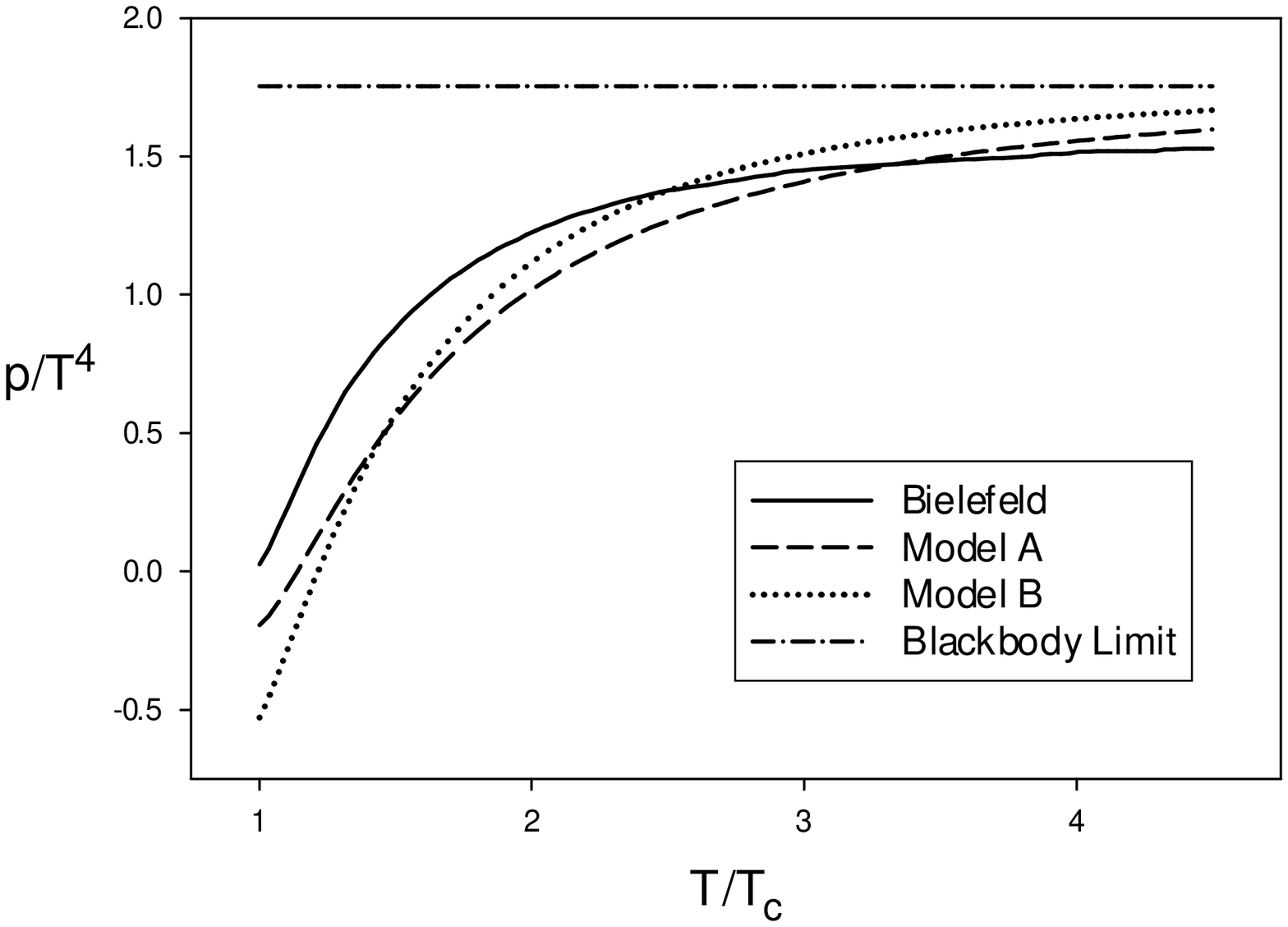}
\caption{$p/T^4$ versus $T/T_c$ for $SU(3)$.}
\label{fig:largenenough}
\end{minipage}
\hspace{\fill}
\begin{minipage}[t]{75mm}
\includegraphics[width=3in]{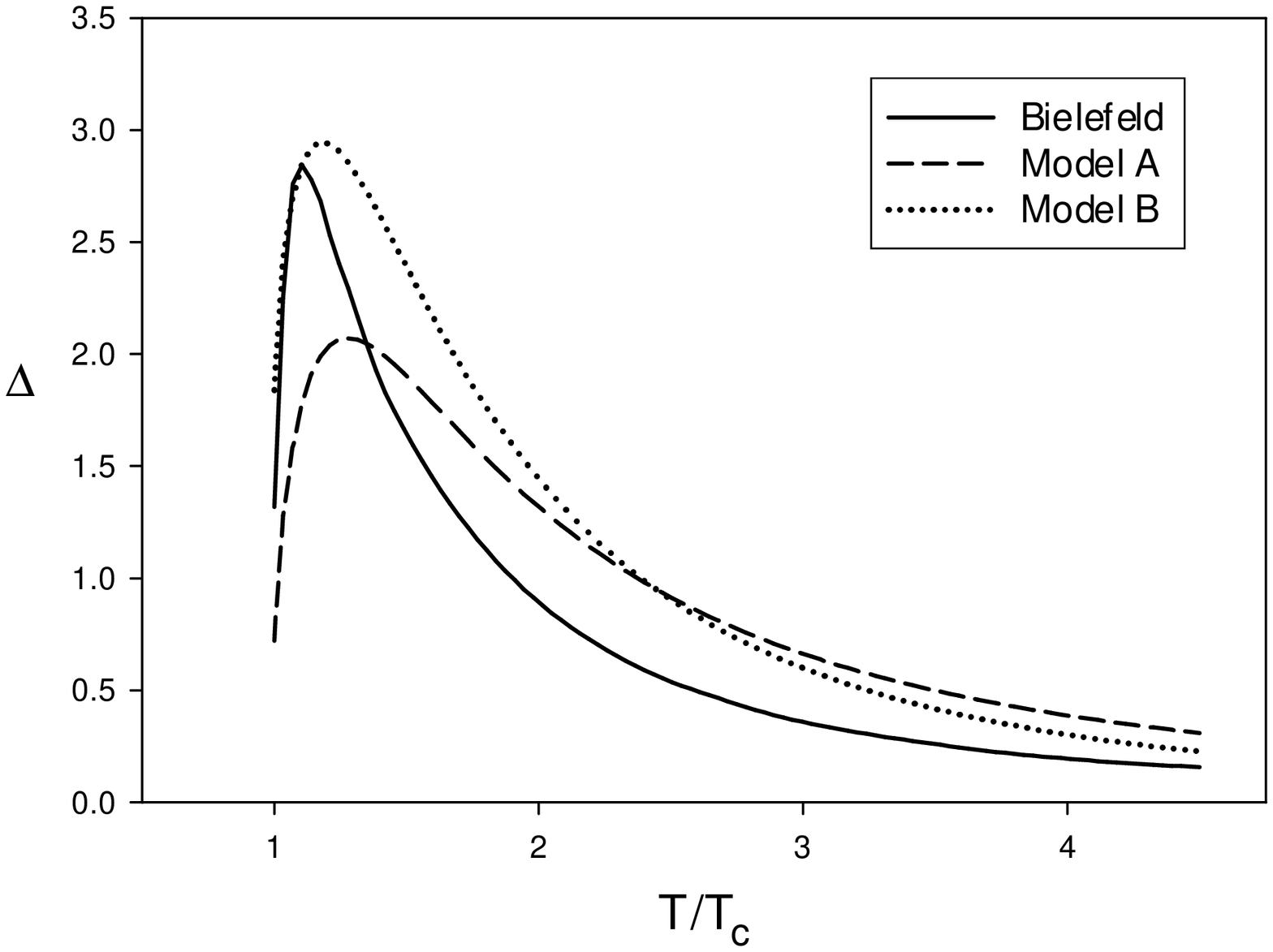}
\caption{$\Delta$ versus $T/T_c$ for $SU(3)$.}
\label{fig:toosmall}
\end{minipage}
\vspace{-0.2in}
\end{figure}

\end{document}